\documentclass[preprint,superscriptaddress]{revtex4-1}

\usepackage{amssymb, amsmath}

\usepackage{float}

\usepackage{graphicx}
\usepackage{color}

\begin{document}

\title{ Magnetotransport of single crystalline YSb}

\author{N. J. Ghimire}
\email{nghimire@anl.gov}
\noaffiliation

\author{A. S. Botana}
\noaffiliation
\author{D. Phelan}
\noaffiliation
\author{H. Zheng}
\noaffiliation
\author{J. F. Mitchell}
\noaffiliation
\affiliation{Materials Science Division, Argonne National Laboratory, 9700 South Cass Avenue, Argonne, Illinois 60439, United States}

\date{\today}

\begin{abstract}
We report magnetic field dependent transport measurements on a single crystal of cubic YSb together with first principles calculations of its electronic structure. The transverse magnetoresistance does not saturate up to 9 T and attains a value of 75,000 \% at 1.8 K. The Hall coefficient is electron-like at high temperature, changes sign to hole-like between 110 and 50 K, and again becomes electron-like below 50 K. First principles calculations show that YSb is a compensated semimetal with a qualitatively similar electronic structure to that of isostructural LaSb and LaBi, but with larger Fermi surface volume. The measured electron carrier density and Hall mobility calculated at 1.8 K, based on a single band approximation, are 6.5$\times10^{20}/$cm$^{3}$ and 6.2$\times10^{4}$cm$^{2}$/Vs, respectively. These values are comparable with those reported for LaBi and LaSb. Like LaBi and LaSb, YSb undergoes a magnetic field-induced metal-insulator-like transition below a characteristic temperature T$_{m}$, with resistivity saturation below 13 K. Thickness dependent electrical resistance measurements show a deviation of the resistance behavior from that expected for a normal metal; however, they do not unambiguously establish surface conduction as the mechanism for the resistivity plateau.
 \end{abstract}
 

\maketitle

\section{Introduction}
The past few years have witnessed considerable effort in the research of  topological states in condensed matter because of the potential for dissipationless flow of electrical current, with ramifications for both energy and electronics \cite{Hasan2010,Hasan2015,Hasan2014,Yang2014,Ando2013,Qi2011,Dzero2010,Moore2010, Xu2015, Yong2012, Xu2015TaAs,Kim2014}. Topological insulators (TIs), have an insulating bulk, but  a conducting surface. The surface states of TIs, which are protected by time reversal symmetry, \cite{Qi2011,Hasan2010} manifest in transport measurements in the form of a resistivity plateau, as observed in a wide range of compounds from Bi$_{2}$Te$_{2}$Se to SmB$_{6}$ \cite{Ren2010, Kim2013, Kim2014, Dzero2010}. Gapless surface states have been recently observed in  Dirac and Weyl semimetals \cite{Xu2015TaAs,Huang2015,Xu2015,Lv2015,Borisenko2014}. These semimetals are materials where conduction and valence bands contact only at some discrete points in the bulk Brillouin zone and show linear dispersion in the vicinity of those points. The surface states of the semimetals are thus distinct from those in the TIs. The surface states of Dirac semimetals are a pair of Fermi arcs that connect the bulk bands at the energy of the bulk Dirac nodes, the point where two linearly dispersing bands are degenerate \cite{Yang2014,Yong2012}. The presence of either (but not both) time reversal symmetry (TR) or  space inversion symmetry (I) lifts the band degeneracy giving rise to a pair of Weyl points \cite{Huang2015} in Weyl semimetals. A distinct transport signature of these gapless surface states of Dirac and Weyl semimetals is difficult to observe, as the transport properties are dominated by the bulk semimetallic behavior. However, quantum oscillations and large and linear  magnetoresistance have been identified as characteristic fingerprints \cite{Huang2015exp,Shekhar2015,Ghimire2015}.

 Recently, behavior reminiscent of the surface states of topological insulators has been realized in a few sets of materials, viz. La$X$ ($X$= Sb, Bi), \cite{Tafti2015,Sun2016,Kumar2016} and T$X_{2}$  (T = Ta, Nb, $X$ = As, Sb) \cite{Wang2014,Luo2016,Li2016,Wu2016}. In the absence of an external magnetic field, the resistivity decreases monotonically with decreasing temperature, indicating that the bulk is gapless. When an external magnetic field is applied, however, the resistivity shows a pronounced upturn, manifesting a metal-insulator-like transition. Interestingly, at low temperatures the resistivity saturates to a constant value and shows a plateau similar to that of the topological insulators. These materials also show a large magnetoresistance and quantum oscillations as in  the compensated  semimetal WTe$_{2}$ \cite{Ali2014} and the Weyl semimetals T$X$ (T = Ta, Nb and $X$ = P, As) \cite{Huang2015exp,Shekhar2015,Ghimire2015}, but the magnetoresistance does not show linear behavior such as that found in Weyl and Dirac semimetals \cite{Liang2015,Xiong2015}.  These new compounds are centrosymmetric. They are also non-magnetic and thus, in zero field, also respect time reversal symmetry. An analysis of LaBi behavior has pointed towards an electron-hole compensation phenomena to be a mechanism for the extremely large magnetoresistance and the upturn of the resistivity in the magnetic field \cite{Sun2016}. However, the origin of the resistivity plateau is still not clear. Here we present a set of magnetotransport data in YSb together with the results of first principles calculations. YSb has similar electronic structure and transport characteristics as those of LaSb and LaBi. Below the metal-insulator-like transition onset-temperature, the electrical resistance of YSb does not scale as the inverse of its thickness as would be expected for a normal metal. However, the low temperature behavior of the thickness-dependent resistance ratio does not support the assignment of the conduction in the region of resistivity plateau solely to the surface states as in the case of the topological insulators.
 
 \section{Experimental details}
\begin{figure}[H]
\begin{center}
\includegraphics[scale=.5]{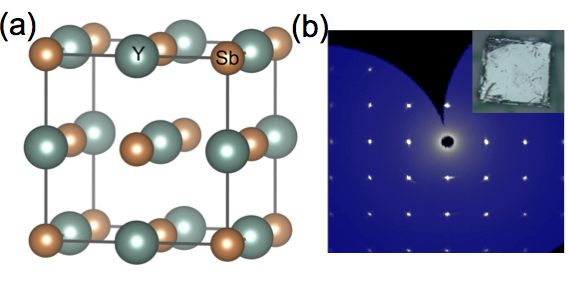}
\caption{a) Graphical sketch of the crystal structure of YSb. b) Single crystal X-ray diffraction pattern of the (100) face of YSb. Optical image showing a cleaved surface of YSb single crystal with dimensions of 1.5 mm $\times$ 1.3 mm is shown on the top right corner.}\label{F0}
\end{center}
\end{figure}

Single crystals of YSb were synthesized in Sb self flux. Y and Sb pieces were loaded in a 2 ml aluminum oxide crucible in a molar ratio of 1:15. The crucible was then sealed in a fused silica ampule under vacuum. The sealed ampoule was heated to 1175$^{\circ}$C over 10 hours, homogenized at 1175$^{\circ}$C for 12 hours and then cooled to 800$^{\circ}$C at the rate of 2$^{\circ}$C/hour. Once the furnace reached 800$^{\circ}$C, the excess flux was decanted from the crystals using a centrifuge. Well faceted cubic crystals as large as 3 mm $\times$ 3 mm $\times$ 3 mm were obtained. An optical image of a cleaved face of a 1.5 mm by 1.3 mm crystal  is shown in the inset of Fig. \ref{F0}(b).

The crystal structure of the compounds was verified by single crystal x-ray diffraction at room temperature using a STOE IPDS 2T diffractometer using Mo K$_{\alpha}$ radiation ($\lambda$ = 0.71073 $\AA$ ) and operating at 50 kV and 40 mA. Integration and numerical absorption corrections were performed using the X-AREA, X-RED, and X-SHAPE programs. Structures were solved by direct methods and refined by full-matrix least squares on F$^{2}$ using the SHELXTL program package \cite{Shelxtl1997}.

Electrical resistivity  measurements were made on an oriented crystal in a Quantum Design Physical Property Measurement System (PPMS) down to 1.8 K and in magnetic fields up to 9 T.  A four wire configuration with 25 $\mu$m gold wire and Epotek H20E silver epoxy were used for the resistivity and magnetoresistance measurements. The Hall effect was measured in a Hall bar geometry. Longitudinal (transverse) voltages were symmetrized (antisymmetrized) about H = 0 T, to account for the small misalignment of the voltage leads.

\section{Results and Discussion}

 The crystal structure of YSb determined from single crystal x-ray diffraction at room temperature is consistent with the NaCl type structure  in cubic space group {\em Fm$\bar{3}$m} (\# 225) as reported in literature \cite{abdusalyamova1990}. The crystallographic data are presented in Table \ref{T1}. Y and Sb reside on  the Wyckoff positions 4b and 4a, respectively. The atomic coordinates of Y are (1/2, 1/2, 1/2) and that of Sb are (0, 0, 0). The lattice parameter obtained from single crystal diffraction is a = 6.1652(7) \AA .  A sketch of the crystal structure of YSb is depicted in Fig. \ref{F0}(a).

\begin{figure}[H]
\begin{center}
\includegraphics[scale=1.3]{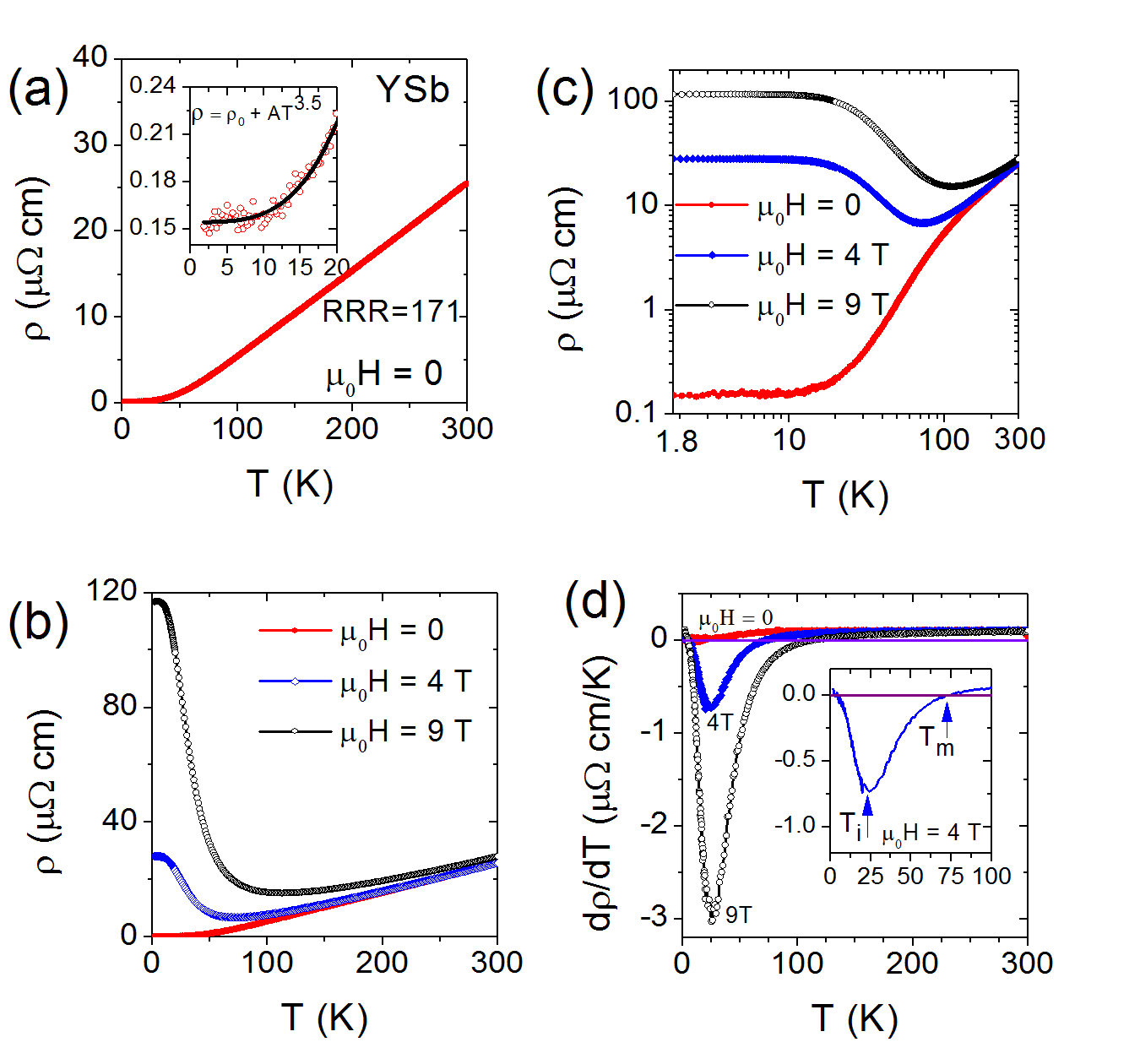}
\caption{ a) Electrical resistivity of YSb as a function of temperature in zero external magnetic field. 
The solid black curve in the inset is a  $\rho$ = $\rho_{0}$ + AT$^{n}$ fit to the resistivity measured below 20 K. b) Electrical resistivity of YSb as a function of temperature in 0, 4 and 9 T with {\bf H} $\bot$ {\bf I} and {\bf H} $\|$ {\em c}-axis. c) Temperature dependence of the electrical resistivity measured in 0, 4 and 9 T shown on a log-log scale. d) Temperature derivative of the electrical resistivity (d$\rho/$dT) of YSb. The peak in d$\rho/$dT marks the inflection point at T$_{i}$. The sign change in d$\rho/$dT marks the resistivity minimum at T$_{m}$.}\label{F1}
\end{center}
\end{figure}

The zero field  resistivity of YSb is shown in Fig. \ref{F1}(a). The resistivity decreases with decreasing temperature from 300 to 13 K,  consistent with a metallic behavior. However, the resistivity levels off below 13 K down to the lowest measured temperature (1.8 K). The residual resistivity ratio (RRR = $\rho$(300 K)/$\rho$(1.8 K) is 171, which reflects a high crystal quality. Below 20 K, the resistivity  fits to $\rho$(T, 0) = $\rho_{0}$ + AT$^{n}$ with $\rho_{0}$ = 0.1545(9) $\mu\Omega$ cm and n = 3.5(2), as shown in the inset of Fig. \ref{F1}(a), a behavior similar to that of LaSb and LaBi \cite{Tafti2015,Sun2016}. Fig. \ref{F1}(b) shows the temperature dependence of the resistivity of YSb measured in transverse magnetic fields of 0, 4 and 9 T applied along the {\em c}-axis. The effect of the magnetic field becomes significant at low temperatures. The resistivity starts to increase with decreasing temperature below 105 K at 9 T and below 70 K at 4 T.  In all  fields, the resistivity saturates  below 13 K  giving a pronounced resistivity plateau, reminiscent of the behavior of the topological insulators attributed to surface states. The low temperature behavior is shown more clearly in a log-log plot of  $\rho$ vs $ T$ in Fig. \ref{F1}(c). The temperature derivative of the resistivity measured at 0, 4 and 9 T is depicted in Fig. \ref{F1}(d).  Similar to LaSb and LaBi \cite{Tafti2015,Sun2016}, a sharp peak is observed at 25 K in both fields, the temperature defined as the inflection point T$_{i}$. At higher temperature $d\rho/dT$ changes sign at T$_{m}$, which is magnetic field dependent: T$_{m}$ at 4 and 9 T is 70 and 105 K, respectively.

\begin{figure}[H]
\begin{center}
\includegraphics[scale=1.3]{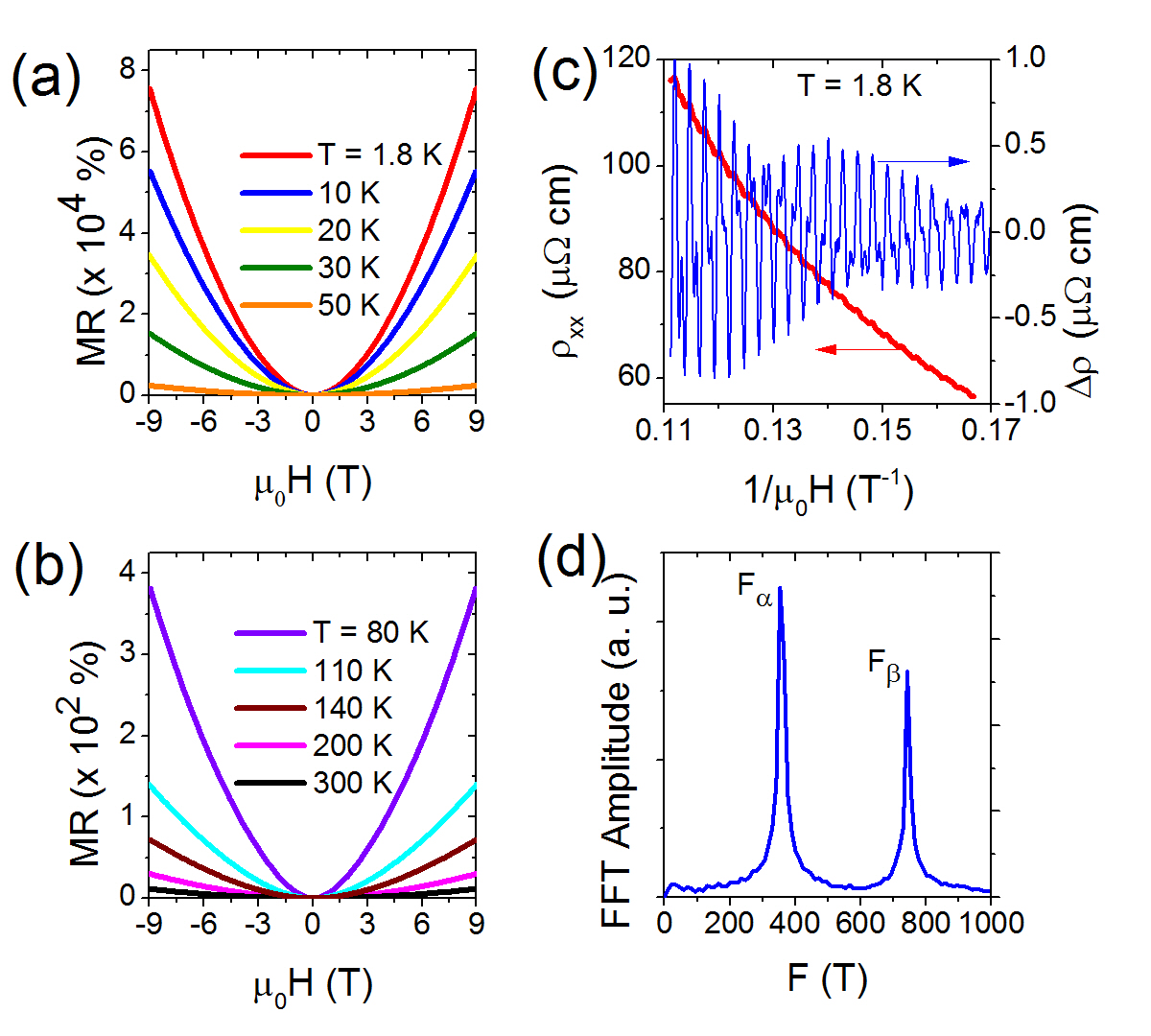}
\caption{a, b) Transverse magnetoresistance of YSb at temperatures indicated.  c) Quantum oscillations observed at 1.8 K between 5 and 9 T (blue curve plotted on the right hand axis). The red curve plotted on the left axis is the MR measured between 5 and 9 T as a function of inverse magnetic field. d) Amplitude of Fast Fourier Transform (FFT) of the data in panel c showing F$_{\alpha}$ = 352 T and F$_{\beta}$ = 740 T.}\label{F2}
\end{center}
\end{figure}

Transverse magnetoresistance, defined as MR = 100 $\times$ [$\rho_{xx}(\mu_{0}H)$ - $\rho_{xx}(\mu_{0}H=0)$]/$\rho_{xx}(\mu_{0}H=0)$ with {\bf H} $\bot$ {\bf I}, is presented in Fig. \ref{F2}(a,b). At room temperature, there is a small MR of about 11 \%. It increases rapidly with decreasing temperature and reaches over 75,000 \% at 1.8 K and 9 T and remains unsaturated. This value compares to that of LaBi \cite{Sun2016} and is about an order of magnitude smaller than that reported for LaSb \cite{Tafti2015,Sun2016}. The magnetoresistance shows a quadratic behavior over the entire temperature range. In addition to the large MR, Shubnikov-de Haas (SdH) oscillations become clear at lower temperatures in the region of the resistivity plateau. The MR above 5 T measured at 1.8 K is shown by the red curve in Fig. \ref{F2}(c). The SdH oscillations $\triangle\rho_{xx}$ = $\rho_{xx}$ - $<$$\rho_{xx}$$>$, obtained  by subtracting a non-oscillatory background $<$$\rho_{xx}$$>$ (a fourth order polynomial fit of $\rho_{xx}$)  from the oscillating $\rho_{xx}$,  is plotted as the blue curve in Fig. \ref{F2}(c). The frequencies of the quantum oscillations obtained by the Fast Fourier Transform (FFT) of $\triangle\rho_{xx}$ as a function of 1/$\mu_{0}$H is depicted in Fig. \ref{F2}(d). Two fundamental frequencies of $F_{\alpha}$ =  352 T and $F_{\beta}$ =  740 T are obtained. These frequencies are larger than in both LaSb and LaBi \cite{Tafti2015,Sun2016} suggesting a slightly larger Fermi surface in YSb than in the other two. This conclusion is supported by the first principles calculations presented below.

\begin{figure}[H]
\begin{center}
\includegraphics[scale=1.3]{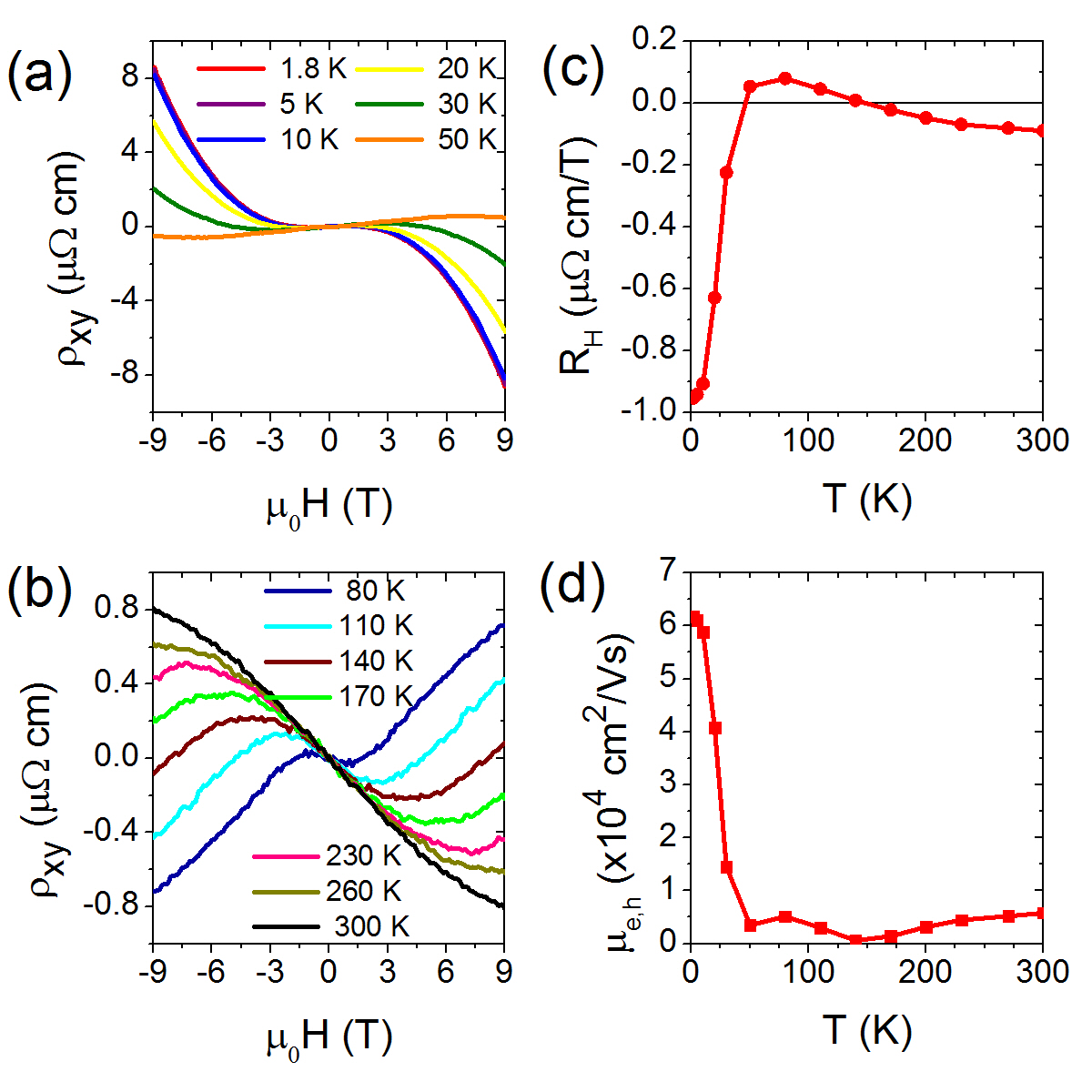}
\caption{ a,b) Hall resistivity of YSb with magnetic field applied along the $a$-axis. c) Hall coefficient R$_{H}$ at 9T as a function of temperature. d) Hall mobility $\mu_{H}$ versus temperature determined by the Hall coefficient at 9 T and zero field resistivity using a single band approximation.}\label{F3}
\end{center}
\end{figure}

The Hall resistivity is defined as $\rho_{xy}$ = ($V_{y}$/$I_{x}$)$t$, where $t$ is the thickness of the sample and $V_{y}$ is the transverse voltage in the presence of a magnetic field applied along $c$-axis with current  $I_{x}$ in the $ab$-plane. The Hall resistivity as a function of magnetic field at different temperatures is plotted in Fig. \ref{F3}(a,b).  At high temperatures, the sign of the Hall resistivity is opposite to that of the magnetic field, indicating electron-like carriers. This behavior changes with decreasing temperature, and becomes positive between 110 and 50 K, again becoming negative below 50 K, suggesting that the majority charge carriers change with temperature. A similar behavior is reported in LaSb \cite{Tafti2015}. The carrier concentration and Hall mobility can be analyzed within a single band model, which is most applicable at the lowest temperatures where the electron contribution dominates the Hall resistivity. The carrier concentration is given by $n$ = 1/($eR_{H}$), where $e$ is the charge of the electron and $R_{H}$ is the Hall coefficient defined by $R_{H}$ = $\rho_{xy}$/($\mu_{0}$H). The temperature dependence of the Hall coefficient taken at 9 T is depicted in Fig. \ref{F3}(c).  The Hall mobility $\mu_{H}$ = 1/en$\rho_{xx} (\mu_{0}H=0)$ is shown in Fig. \ref{F3}(d). At 1.8 K, the computed carrier concentration is 6.5$\times 10^{20}/$cm$^{3}$, which is an upper bound since the hole contribution is not considered under the single band approximation. The Hall mobility at 1.8 K is 6.2$\times10^{4}$ cm$^{2}$/Vs. Again it is similar to LaBi \cite{Sun2016} and about an order of magnitude smaller than that of LaSb \cite{Tafti2015}.

\begin{figure}[H]
\begin{center}
\includegraphics[scale=1.5]{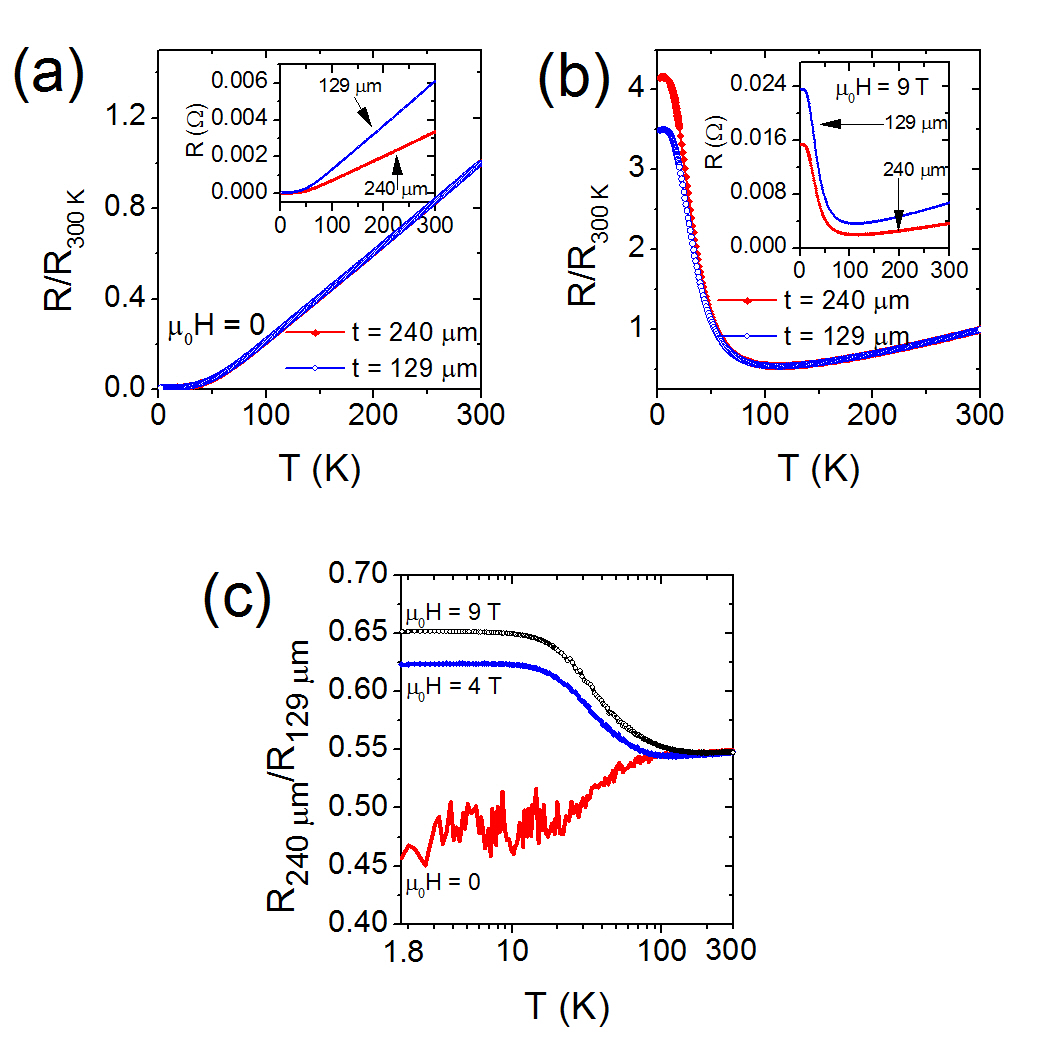}
\caption{Temperature dependence of normalized resistance R(T)/R(300 K) of a sample with thicknesses of 240 and 129 $\mu$m in an external magnetic field of a) 0 and b) 9 T. The insets show temperature dependence of the bare resistance. c) Ratio of resistance of 129 $\mu$m  sample to that of  240 $\mu$m sample as a function of temperature. The electrical contacts attached to the  240 $\mu$m sample before polishing were kept intact in the 129 $\mu$m sample.}\label{F4}
\end{center}
\end{figure}

From Ohm's  law, the electrical resistance of a sample is given by R = $\rho l/A$, where $\rho$ is the resistivity of the material, $l$ is the length and $A$ is the cross-sectional area. If the thickness of a uniform metallic rectangular bar is decreased by a factor $k$, the electrical resistance should increase by the same factor. We tested this for YSb. First, the electrical resistance of a 240 $\mu$m thick sample was measured. Four gold stripes were made on a surface of the sample by gold sputtering to ensure the better contact between the gold wires and the sample. Gold wires were fixed on top of the stripes using silver epoxy (Epotek H20E). After the first measurement, the sample was polished on the back side to a thickness of 129 $\mu$m. All four gold wires were kept intact. We found that for T $\gtrsim$150 K the ratio of the resistance measured on the sample before and after polishing is equal to the ratio of the two thicknesses ($k$=0.54) [Fig. \ref{F4}(c)], which guarantees that the only change in the sample before and after polishing was its thickness. Figure \ref{F4}(a) and \ref{F4}(b) show the temperature dependence of normalized resistance R(T)/R(300 K)  of the 240 and 129 $\mu$m samples, measured in an external magnetic field of 0 and 9 T, respectively. The bare resistances are shown in the insets. The bare resistance of the thinner sample is larger both in 0 and 9 T  over the entire temperature range. However, the normalized resistances of the two samples overlap at high temperature and change as the temperature decreases. A ratio of the resistance of 129 and 240 $\mu$m samples (R$_{129}$/R$_{240}$) in 0, 4 and 9 T is depicted in Fig. \ref{F4}(c). It is seen that this ratio is constant and equal to the ratio of the thickness above T$_{m}$. Below T$_{m}$, this ratio changes and again becomes constant below 13 K, where the resistivity plateau appears, indicating that in YSb the resistance does not scale to the inverse of the cross-sectional area as expected for a normal metal,  irrespective of the presence of magnetic field.  Deviation of such a scaling has also been observed in the topological Kondo insulator SmB$_{6}$, \cite{Kim2014} where the behavior is attributed to a conducting surface state. In SmB$_{6}$, the ratio of the resistance attains the value of unity providing evidence that the conduction occurs only on the surface. This ratio in YSb fails to reach unity even at the field of 9T, which suggests that the conduction channel in YSb up to 9 T is not solely through the surface.  Interestingly, although the metal-insulator-like transition in YSb is induced by external magnetic field, the deviation of the scaling behavior is evident even in zero field [Fig. \ref{F4}(c)].

First principles calculations have been carried out to shed light into the striking similarity between the transport properties of YSb and those of LaSb and LaBi. The electronic structure calculations have been carried out within density functional theory (DFT) using the all-electron, full potential code WIEN2K \cite{wien2k} based on the augmented plane wave plus local orbital (APW + lo) basis set \cite{APWlo}. The Perdew-Burke-Ernzerhof (PBE) version of the generalized gradient approximation(GGA) \cite{GGA} was chosen as the exchange correlation potential. Spin orbit coupling (SOC) was introduced in a second variational procedure \cite{SOC}. A dense k-mesh  of  $39\times39\times39$ was used for the Brillouin Zone (BZ) sampling in order to check the fine details of the influence of spin orbit coupling in the electronic structure. A R$_{mt}$K$_{max}$ of 7.0 was chosen for all the calculations.

\begin{figure}[H]
\begin{center}
\includegraphics[scale=.5]{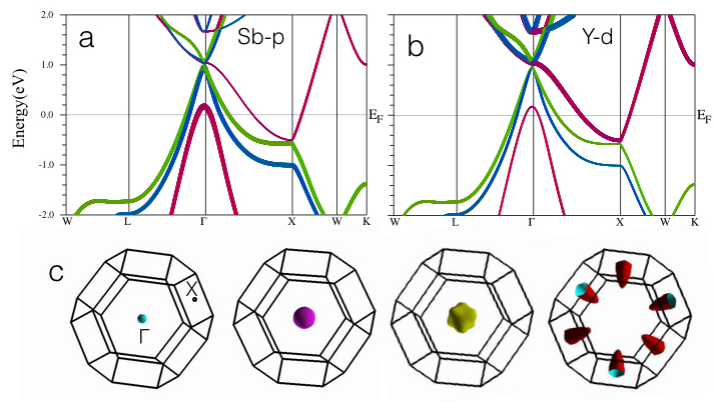}
\caption{ Band structures with band character plot of YSb along high symmetry directions of the Brillouin zone highlighting (a) Sb-p  character and  (b) Y-d character, and (c) Fermi surface of YSb obtained within GGA+SOC showing three nearly spherical hole pockets at the $\Gamma$ point and ellipsoidal electron pockets centered at the X point.}\label{F5}
\end{center}
\end{figure}

Figures \ref{F5}(a, b) show the band structure with band character plot along high symmetry directions of the Brillouin zone with inclusion of SOC. The electronic structure corresponds to that of a compensated semimetal, with overlap of conduction and valence bands resembling that of LaSb and LaBi \cite{Gou2016}. The valence band has mainly Sb-$p$ character whereas the conduction band has primarily Y-$d$ character. The Fermi surfaces of YSb calculated with the inclusion of SOC are  shown in Fig. \ref{F5}(c). There are three nearly spherical hole pockets around the $\Gamma$ point  with the equivalent ellipsoidal electron pockets centered at the X point. The electron and hole pockets have a larger volume in YSb  as compared to LaSb and LaBi, and this is consistent with the experimental finding that the quantum oscillation frequencies are larger in YSb than in the other two compounds. Our calculations indicate that the resemblance in the transport properties of YSb and LaSb/LaBi stems from their similar electronic structures, which comprise compensating electron/hole carriers at the $\Gamma$ and X points of the Brillouin zone.

\section{Conclusion}
In summary, we have grown single crystals of YSb and studied its electrical transport properties.  Overall, YSb behaves similarly to previously reported studies of LaSb and LaBi \cite{Tafti2015,Sun2016}. The magnetotransport displays a pronounced magnetoresistance, which does not saturate up to 9 T, with a large Hall mobility of 6.2$\times10^{4}$ cm$^{2}$/Vs at 1.8 K. The electronic structure of YSb corresponds to that of a compensated semimetal and is qualitatively similar to that of LaSb and LaBi. The results obtained from a thickness-dependent study indicate that the resistivity plateau observed at least up to 9 T cannot exclusively be attributed to surface conduction. The large magnetoresistance and the resistivity upturn in the external magnetic field below T$_{m}$ have been attributed to the electron-hole compensation mechanism in LaBi \cite{Sun2016} and, possibly, in  LaSb, \cite{Tafti2015} and YSb. It is interesting that the electrical resistance of YSb does not scale to the inverse of the thickness over the same temperature region. The centrosymmetric crystal lattice in this non-magnetic compound does not support the formation of the Fermi arc as in the Weyl semimetals. Likewise, quadratic behavior of the MR in all the measured temperatures is different from the linear MR behavior in Dirac semimetals Na$_{3}$Bi and Cd$_{3}$As$_{2}$ \cite{Liang2015,Xiong2015}. Further investigation by, for example, Angle-Resolved Photoemission may shed further light on this issue. YSb provides an attractive system for such investigations because large single crystals with dimensions up to 3 mm are easily grown out of the Sb self flux, a consequence of the wide stability of YSb in an accessible temperature range - between 750 and 1200 $^{\circ}$C in the Y-Sb phase diagram \cite{YSbPhasediagram1990}.

\section{Acknowledgements}
This work was supported by the U.S. Department of Energy, Office of Science, Basic Energy Sciences, Materials Science and Engineering Division. The authors thank Duck Young Chung for his help during the crystal growth.
\newpage

\begin{table}[H]
\caption{Crystallographic data and Atomic Coordinates and Equivalent Displacement Parameters for YSb. U$_{eq}$ is defined as one third of the trace of the orthogonalized U$^{ij}$ tensor.  $\hspace{2cm}$}\label{T1}
\begin{tabular*}{1\linewidth}{@{\extracolsep{\fill}}ll}
\hline
   Crystal system            &                                 Cubic	$\hspace{10.0cm}$                                                                                                    \\
   Space group               &         	                       {\em Fm$\bar{3}$m}	                                                                                                \\
   Temperature (K)              &                                  293(2)                                                                                                                            \\
   Wavelength (\AA)                &                                 0.71073                                                                                                       \\
   Z formula units           &                                 4                                                                                                               \\
   2$\theta_{min}$          &                                5.73$^{o}$                                                                                                         \\
   2$\theta_{max}$         &                               28.38$^{o}$                                                                                                       \\		
   Formula weight (g/mol)           &                              210.66  \\
  a (\AA)                         &	                            6.1652(7) 	    	       \\
  Volume (\AA$^{3}$)	      &	                            234.34(8) 	           \\
  Density (calculated) (g/cm$^{3}$)       &	  5.971  	              \\
  $\mu$ (Mo K$\alpha$) (cm$^{-1}$)   &      357.09                 \\
  Goodness-of-fit on F$^{2}$               &	1.144  	                 \\
  $R(F)$ for $F_{o}^{2}$ $>$               &                          \\
  2$\sigma (F_{o}^{2})^{a}$                 &    0.0147                  \\
  $R_{w}(F_{o}^{2})^{b}$                    &   0.0358	               \\
\end{tabular*}
\begin{tabular}{l@{\hspace{1.3cm}}l@{\hspace{1.3cm}}l@{\hspace{1.3cm}}l@{\hspace{1.3cm}}l@{\hspace{1.3cm}}l@{\hspace{1.3cm}}l}				\hline																		
        Atom           & Wyck.      & Occ.   &	  x	          &	      y	         &	  z	        &   U$_{eq}$ (\AA$^{2}$)   \\
\hline													
         Y	       &	4b	    &	1	&	1/2	          &	  1/2	             &	  1/2	        &	 0.012(1)	\\
         Sb	       &	4a	    &	1	&	0	          &	   0	             &	   0       	&	0.014(1)	\\   					
\hline
\end{tabular}

\begin{tabular*}{1\linewidth}{ll}
 $^{a}R(F)$ = $\sum\mid\mid F_{o}\mid -\mid F_{c}\mid\mid/\sum\mid F{_o}\mid$ &  $^{b}R_{w}(F_{o}^{2})$	= $[\sum[w(F_{o}^{2}-F_{c}^{2})^{2}]/\sum wF_{o}^{4}]^{1/2}$   \\ 	 \end{tabular*}				
\end{table}

\bibliographystyle{ieeetr}

\end{document}